\documentclass[preprint,12pt,authoryear]{elsarticle}

\usepackage{amsmath, amssymb, verbatim, subcaption, bm,  mathrsfs}
\usepackage{hyperref}
\usepackage[frozencache,cachedir=.]{minted}
\usepackage{pifont}
\newcommand{\cmark}{\ding{51}}%
\newcommand{\xmark}{\ding{55}}%
\usepackage{makecell}

\newcommand{\proglang}[1]{\textsf{#1}}
\newcommand{\pkg}[1]{\texttt{#1}}
\usepackage{amsthm}
\usepackage{enumitem}
\usepackage{float}
\newtheorem{theorem}{Theorem}
\newtheorem{definition}[theorem]{Definition}
\newtheorem{example}[theorem]{Example}

\usepackage[linesnumbered , ruled, noend, norelsize]{algorithm2e}
  \SetKwInOut{Input}{Input}
  \SetKwInOut{Output}{Output}
\IncMargin{1em}

\usepackage{xcolor}

\usepackage{tikz}
\usetikzlibrary{shapes,arrows}

\newcommand{\calT}{\mathcal{T}}
\newcommand{\calC}{\mathcal{C}}
\newcommand{\calS}{\mathcal{S}}
\newcommand{\scrE}{\mathscr{E}}

\journal{Knowledge-Based Systems}

\begin{document}

\begin{frontmatter}



\title{cegpy: Modelling with Chain Event Graphs in Python}


\author[inst1]{Gareth Walley\fnref{label1}}

\author[inst2]{Aditi Shenvi\corref{cor1}\fnref{label1}}
\ead{aditi.shenvi@gmail.com}

\author[inst3,inst4]{Peter Strong}

\author[inst2]{Katarzyna Kobalczyk}

\cortext[cor1]{Corresponding author.}

\fntext[label1]{Both authors contributed equally to this research.}

\affiliation[inst1]{
            city={Kenilworth},
            postcode={CV8 1JY}, 
            country={UK}}
\affiliation[inst2]{organization={Statistics Department, University of Warwick},
            city={Coventry},
            postcode={CV4 7AL}, 
            country={UK}}
\affiliation[inst3]{organization={Centre for Complexity Science, University of Warwick},
            city={Coventry},
            postcode={CV4 7AL}, 
            country={UK}}
            
\affiliation[inst4]{organization={Alan Turing Institute},
            city={London},
            postcode={NW1 2DB}, 
            country={UK}}
            
\begin{abstract}
Chain event graphs (CEGs) are a recent family of probabilistic graphical models that generalise the popular Bayesian networks (BNs) family. Crucially, unlike BNs, a CEG is able to embed, within its graph and its statistical model, asymmetries exhibited by a process. These asymmetries might be in the conditional independence relationships or in the structure of the graph and its underlying event space. Structural asymmetries are common in many domains, and can occur naturally (e.g. a defendant vs prosecutor's version of events) or by design (e.g. a public health intervention). However, there currently exists no software that allows a user to leverage the theoretical developments of the CEG model family in modelling processes with structural asymmetries. This paper introduces \pkg{cegpy}, the first \proglang{Python} package for learning and analysing complex processes using CEGs. The key feature of \pkg{cegpy} is that it is the first CEG package in any programming language that can model processes with symmetric as well as asymmetric structures. \pkg{cegpy} contains an implementation of Bayesian model selection and probability propagation algorithms for CEGs. We illustrate the functionality of \pkg{cegpy} using a structurally asymmetric dataset.
\end{abstract}




\end{frontmatter}



\section{Introduction} \label{sec:intro}

A probabilistic graphical model (PGM) is composed of a statistical model and a graph representing the conditional independence relationships between the defining random variables or events of the underlying model. The graph of a PGM gives a compact visual representation of the factorisation of the joint probability distribution of a statistical model, and provides a way to perform efficient inference using local computations \citep{pearl2009causality}. A key benefit of PGMs is that the gist of its graph can typically be understood by those without formal statistical or mathematical training; thereby facilitating interactions between statisticians, domain experts and decision makers. 

Bayesian networks (BNs) \citep{pearl2009causality}, currently the most popular family of PGMs, have been successfully applied to a wide range of domains. Notwithstanding the great success of BNs, they do have some shortcomings. In particular, BNs are unable to fully describe processes that exhibit asymmetries either in their conditional independence relationships or in their structure. The former indicates the presence of context-specific conditional independencies which are independence relationships that only hold for certain values of the conditioning variables. Whereas, the latter refers to the presence of structural missing values, i.e. values that are missing which have no underlying meaningful value, and/or structural zeros\footnote{These are in contrast to sampling zeros which occur due to limitations of data sampling.}, i.e. observations of a zero frequency for a category of a categorical variable where a non-zero observation is logically restricted. Such asymmetries are common in many real-world processes especially in domains such as medicine, risk analysis, policing, forensics, law, ecology and reliability engineering where processes are best described through an unfolding of events (see e.g.  \citet{zhang1999role, boutilier1996context, shenvi2018modelling}).

Chain event graphs were introduced in \citet{smith2008conditional} for asymmetric processes and they generalise BNs. A CEG is constructed from the event tree description of a process by leveraging the probabilistic symmetries existing within it to reduce the number of nodes and edges required. Thereby, CEGs ensure the resultant statistical model and graph are not any more complex than they need to be to represent the process. Since their relatively recent introduction, several methodological developments have now been made for the CEG family. These include model selection algorithms \citep{freeman2011bayesian, silander2013dynamic, cowell2014causal}, probability propagation algorithms \citep{thwaites2008propagation, shenvi2020propagation} and a d-separation theorem \citep{wilkerson2020thesis} as well as tools for causal inference \citep{thwaites2010causal, thwaites2013causal, yu2021causal} and diagnostics in a CEG \citep{wilkerson2020thesis}. Applications of CEGs in public health \citep{barclay2013refining, shenvi2018modelling}, medicine \citep{keeble2017adaptation, keeble2017learning}, educational studies \citep{freeman2011dynamic}, asymmetric Bayesian games \citep{thwaites2017new}, migration studies \citep{strong2021bayesian}, and policing \citep{collazo2017thesis, bunnin2019bayesian} have also been explored.

Despite the prevalence of asymmetric processes and the proven flexibility offered by CEGs in modelling such processes, CEGs are yet to be widely adopted by applied statisticians. A major hindrance to the wider application of CEG methodologies is the lack of existing software, particularly when it comes to modelling structurally asymmetric processes. There currently exist two \proglang{R} packages for modelling with CEGs, namely \pkg{ceg} \citep{ceg} and \pkg{stagedtrees} \citep{carli2022r}. However, neither of these supports processes with structural asymmetries. In contrast, for modelling with BNs, there exist several well-developed and maintained softwares such as Netica \citep{netica}, Weka \citep{eibe2016weka}, BARD \citep{nyberg2022bard}, GeNIe \citep{genie}, and Hugin \citep{hugin}, as well as packages such as \pkg{bnlearn} \citep{bnlearn} and \pkg{deal} \citep{deal} in \proglang{R}; \pkg{BayesPy} \citep{luttinen2016bayespy}, \pkg{GOBNILP} \citep{cussens2020gobnilp} and \pkg{BayeSuites} in \proglang{Python}. In this paper, we present \pkg{cegpy}, the first \proglang{Python} package for modelling with CEGs, and the first package across all languages to support modelling of processes with structural asymmetries. 

Section \ref{sec:modelling_with_cegs} discusses the prevalence of structurally asymmetric processes, and it provides a review of CEGs as is relevant for \pkg{cegpy}. Section \ref{sec:cegpy_package} discusses the implementation of \pkg{cegpy} in \proglang{Python}. Section \ref{sec:worked_example} illustrates the strength of \pkg{cegpy} through modelling of an asymmetric public health intervention designed to reduce falls-related injuries among the elderly. We conclude the paper with a discussion about future functionalities for \pkg{cegpy} in Section \ref{sec:discussion}.

\section{Modelling with Chain Event Graphs} \label{sec:modelling_with_cegs}

\subsection{Asymmetric Processes} \label{subsec:asymmetries}

Before reviewing CEGs, we first discuss the generality of asymmetric processes. In particular, the main advantage of CEGs over other PGMs is that they can model processes with structural asymmetries. Asymmetric structures may occur naturally (e.g. prosecutor vs defendant's versions of events in a courtroom trial) or by design (e.g. an intervention to provide all or some of (i) behavioural counselling, (ii) group therapy and (iii) nicotine replacement products to individuals depending on their smoking history). Below we list further examples of structurally asymmetric processes.


\begin{itemize}[itemsep=0pt, topsep=0.5em]
    \item \textit{A public health intervention process:} To maximise the effects of the intervention given limited resources, the level of support is typically varied based on the needs of the individuals (e.g. annual flu vaccines provided by the UK's National Health Services to those most at risk). 
    \item \textit{A medical diagnosis process:} Diagnosis involves identifying the patient's ailment based on their background covariates (e.g. age, sex, health history), the sequence of symptoms already exhibited by them, and the additional symptoms that they may exhibit in the future according to the possible diagnoses. Different diseases might have an overlap of symptoms, and further, not all patients suffering from a certain disease exhibit all associated symptoms. 
    \item \textit{A forensic proceeding:} The contrasting explanations for forensic evidence given by a defendant and prosecutor typically result in asymmetric developments in the sequence of events in the two arguments.
    %
    \item \textit{A policing process:} Depending on the current preparedness of an individual intent on committing a crime, the sequence of preparatory tasks undertaken by the individual can be very varied; e.g.: training, help of like-minded criminals, target identification acquisition of resources.  
    \item \textit{A machinery failure process:} Several different sequences of events can lead to faults in machinery, involving different failing components. 
    %
\end{itemize}

As evidenced by the above examples, we conjecture that asymmetric processes are common -- or at the very least, not uncommon -- in several domains. However, a large number of statistical methods and techniques rely on using variables as building blocks in their description of the process, and thereby, implicitly expect process to have symmetric structures. This expectation influences the choice of variables used to describe a process (e.g. choosing proxy variables to circumvent structural missing values) and also the way in which the data is recorded. In Section \ref{sec:worked_example} we describe how structural zeros and structural missing values can be easily recorded within the dataset of an asymmetric process for modelling with CEGs within our \pkg{cegpy} package.

\subsection{Chain Event Graphs} \label{subsec:ceg}

CEGs provide an alternative to BNs for processes exhibiting asymmetries in their conditional independencies and/or structures. With major modifications (typically tree-based), a BN can represent context-specific conditional independencies. However, currently there is no way to embed structural asymmetries within a BN model. The existence of asymmetric structures has been recognised in several domains, but until the development of CEGs, had not been explicitly addressed within the graphical modelling framework. 

The construction of the CEG for a process begins by eliciting the event tree describing the process being studied. An event tree is a directed tree graph with a single root node. The nodes with no emanating edges are called leaves, and the non-leaf nodes are called situations. Each situation is associated with a transition parameter vector which indicates the conditional probability of an individual, who has arrived at the situation, going along one of its edges. The event tree undergoes a sequence of transformations to become the graph of its associated CEG. A non-technical summary of these transformations as given in \citet{shenvi2020constructing} is presented below.

\begin{itemize}[itemsep=0pt, topsep=0.5em]
    \item Situations in the event tree whose immediate evolutions, i.e. their associated conditional transition parameter vectors, are equivalent are said to be in the same \textit{stage} and are assigned the same node colouring to indicate this symmetry.
    \item Situations whose rooted subtrees (i.e. the subtree obtained by considering that situation as the root) are isomorphic in the structure and colour preserving sense are said to be in the same \textit{position} and are merged into a single node. This node retains the colouring of the situations it merged. 
    \item All the leaves are merged into a single \textit{sink} node. 
\end{itemize}

Formally, let $\calT$ denote an event tree with a finite node set $V(\calT)$ and an edge set $E(\calT)$. A directed edge $e \in E(\calT)$ from nodes $s_i$ to $s_j$ with label $l$ is an ordered triple given by $(s_i, s_j, l)$; or by $e_{ij}$ when unambiguous. Let $L(\calT)$ and $S(\calT) = V(\calT) \backslash L(\calT)$ denote the sets of leaves and situations respectively. 

\begin{definition}[Stage] 
In an event tree $\calT$, two situations $s_i$ and $s_j$ are said to be in the same stage whenever $\pmb{\theta}_{s_i} = \pmb{\theta}_{s_j}$. Additionally, for $\theta(e_i) = \theta(e_j)$ we require that $e_i = (s_i, \cdot, l)$ and $e_j = (s_j, \cdot, l)$ where edge $e_i$ emanates from $s_i$ and $e_j$ emanates from $s_j$. 
\end{definition}

Identification of the stages in the event tree can be done using any suitable model selection algorithm. The collection of stages $\mathbb{U}$ partitions the nodes of the event tree. An event tree whose nodes are coloured to represent its stage memberships is called a \textit{staged tree}. It is common practice to suppress the colouring of singleton stages in the graphs of the staged tree and CEG for visual clarity. Within \pkg{cegpy}, singleton stages are coloured white, leaves and their corresponding sink node are in light-grey. Stages enable us to reduce the parameter space of the CEG model. For compacting the representation of the CEG model, an additional concept of \textit{positions} is required. 

\begin{definition}[Position]
In a staged tree $\calS$, two situations $s_i$ and $s_j$ are said to be in the same position whenever we have $\Phi_{\calS_{s_i}} = \Phi_{\calS_{s_j}}$ where $\calS_{s_i}$ and $\calS_{s_j}$ are the coloured subtrees of $\calS$ rooted at $s_i$ and $s_j$ respectively.
\end{definition}

The collection of positions $\mathbb{W}$ is a finer partition of the node set of the event tree. Situations which are in the same position can be represented by a single node in the graph of the CEG as their complete future unfoldings are identical. We now formally define a CEG. 

\begin{definition}[Chain Event Graph]
A CEG $\calC = (V(\calC), E(\calC))$ is defined by the triple $(\calS, \mathbb{W}, \Phi_\calS)$ with the following properties:
\begin{itemize}[itemsep=0pt, topsep=0.5em]
    \item $V(\calC) = R(\mathbb{W}) \cup w_\infty$ where $R(\mathbb{W})$ is the set of situations representing each position set in $\mathbb{W}$, $w_\infty$ is the sink node and for $w \in V(\calC)$, $\theta_\calC(w) = \theta_\calS(w)$. Nodes in $R(\mathbb{W})$ retain their stage colouring.
    \item Situations in $\calS$ belonging to the same position set in $\mathbb{W}$ are contracted into their representative node contained in $R(\mathbb{W})$. This node contraction merges multiple edges between two nodes into a single edge only if they share the same edge label.
    \item Leaves of $\calS$ are contracted into sink node $w_\infty$.
\end{itemize}
\end{definition}

Staged trees and CEGs that embed context-specific independencies but not structural asymmetries are said to be \textit{stratified}. Those that additionally embed structurally asymmetries are said to be \textit{non-stratified}.  

\begin{figure}
    \centering
    \begin{subfigure}{0.8\textwidth}
        \centering
        \includegraphics[width=0.7\textwidth]{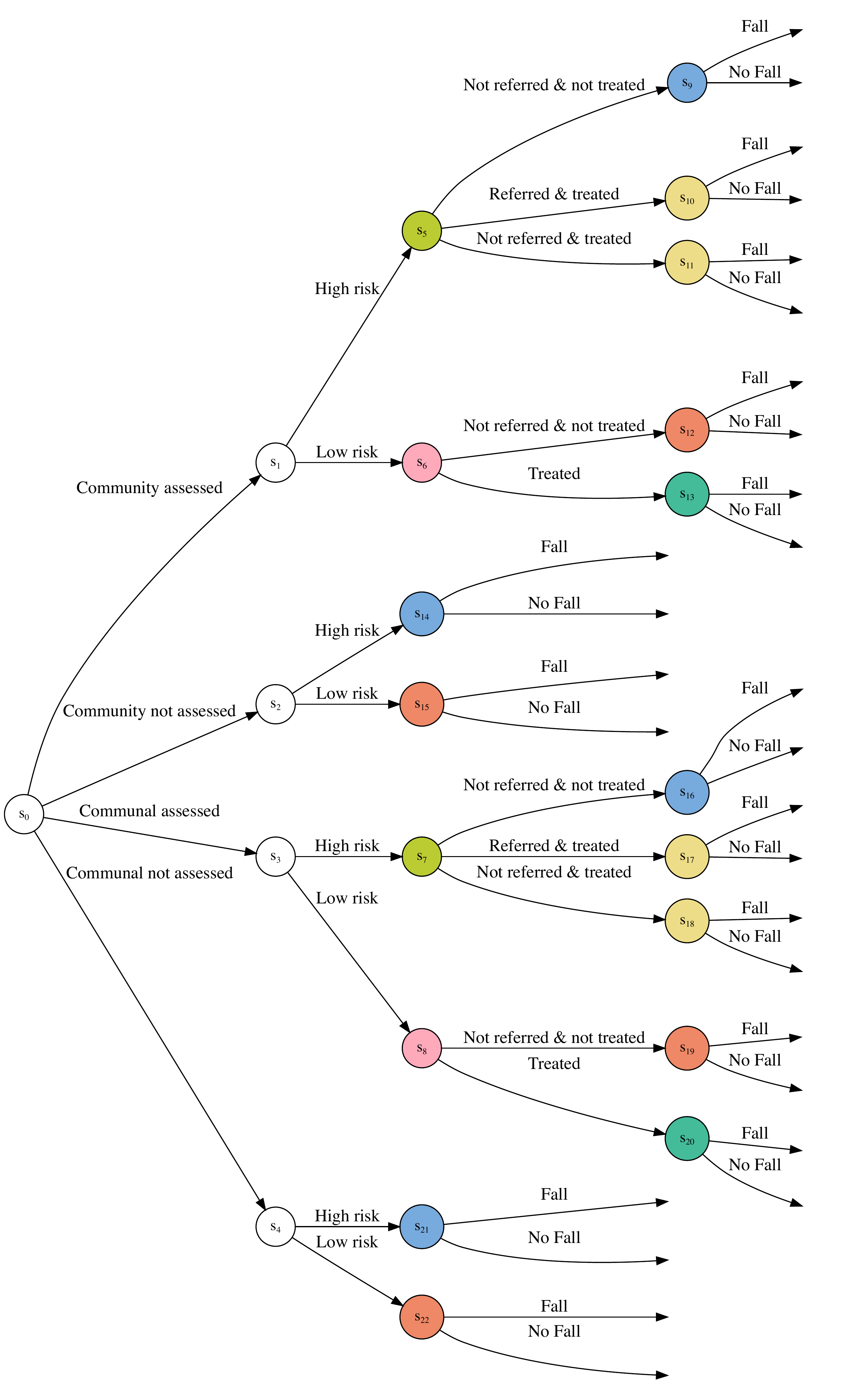}
        \caption{}
    \end{subfigure}
    \begin{subfigure}{0.8\textwidth}
        \centering
        \includegraphics[width = 0.9\textwidth]{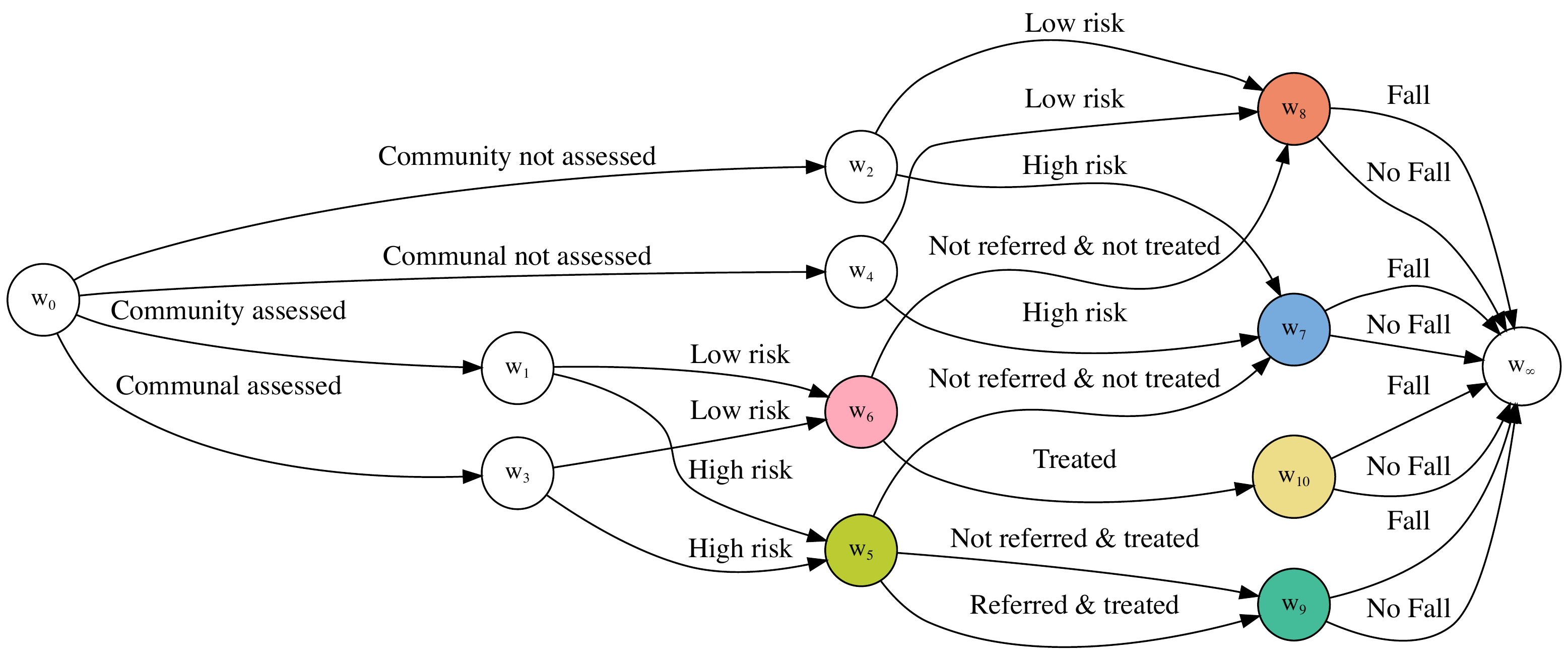}
        \caption{}
    \end{subfigure}
    \caption{(a) A hypothesised staged tree and (b) the corresponding CEG for the falls intervention described in Example \ref{ex:falls_description}.}
    \label{fig:falls_event_staged_tree_ceg}
\end{figure}

\begin{example}[Falls Intervention]
\label{ex:falls_description}
We consider here a non-pharmacological intervention designed to reduce falls-related injuries and fatalities among the elderly as presented in \citet{eldridge2005modelling} and later modelled with a CEG in \citet{shenvi2018modelling}. The intervention aims to enhance assessment, referral pathways, and treatment for individuals aged over 65 years living in the community as well as those in communal establishments (i.e. care homes, nursing homes and hospitals) who have a substantial risk of falling. Under this intervention, a certain proportion of individuals aged over 65 would be assessed and classified as low or high risk. Those assessed to be at a high risk are referred to a falls clinic for an advanced assessment. All those who are referred, 50\% of other high risk individuals, and 10\% of low risk individuals go on to receive treatment. It is assumed that those who are not assessed receive neither referral nor treatment. 

Figure \ref{fig:falls_event_staged_tree_ceg} shows a hypothesised staged tree and its corresponding CEG for this process. It is clear from the staged tree that the process has an asymmetric structure by interventional design, and therefore the staged tree and CEG representing this process are non-stratified. Nodes $s_{5}$ and $s_7$ are coloured the same to indicate that having arrived at one of these nodes, the probability of being referred and treated (/not being referred but being treated/ not being treated) is the same. These nodes are also in the same position as their rooted subtrees are isomorphic in the colour- and structure-preserving sense. Therefore, $s_{5}$ and $s_7$ are represented by the single node $w_5$ in the graph of the CEG.   
\end{example}

\subsubsection{Bayesian Model Selection} \label{subsubsec:model_selection}

A CEG is uniquely defined by its staged tree and the parameters over its staged tree, $\Phi_{\calS}$ \citep{shenvi2020constructing}. Hence, model selection in CEGs is equivalent to identifying the sets of stages in the event tree to obtain the staged tree. The two main approaches to learning the stages in an event tree are (i) the agglomerative hierarchical clustering (AHC) algorithm \citep{freeman2011bayesian} and (ii) a brute-force approach using dynamic programming \citep{cowell2014causal, silander2013dynamic}. Both of these are score-based algorithms that aim to maximise a chosen score function, typically the log marginal likelihood score of the model. 

Within \pkg{cegpy}, we implemented the AHC algorithm as it is computationally efficient for moderate-sized event trees -- which we envision will be what the package will be used for -- and at present, it is the most popular approach for applications involving CEGs. The AHC algorithm does not rely on structural symmetry and is directly applicable to the non-stratified class. 

The AHC algorithm is a greedy, bottom-up hierarchical clustering algorithm. It begins with each situation in the event tree being in its own singleton stage. Thereafter, at each step, it merges the two stages that give the highest improvement in terms of the log marginal likelihood score. The algorithm terminates when the score can no longer be improved by merging two stages. The technical details of the AHC and its associated pseudo-code are presented in Appendix A in the supplementary material. 

\subsubsection{Probability Propagation} \label{subsubsec:probability_propagation}

Probability propagation refers to updating of posterior parameters using local computations, given the observation of some ``evidence". The probability propagation algorithm for CEGs was given by \citet{thwaites2008propagation}. 

Observation of evidence for CEGs corresponds to the observation that an individual (or equivalently, a set of exchangeable individuals) has visited a node or an edge. This type of evidence may be referred to as \textit{positive} evidence where we observe that a node/edge was visited as compared to \textit{negative} evidence where we observe that a node/edge was not visited. Further, evidence is \textit{certain} when observations occur with probability one and \textit{uncertain} evidence when there is a non-trivial probability distribution associated with a possible set of events or states. Note that negative evidence can be recast as uncertain positive evidence. The CEG propagation algorithm requires that the evidence is \textit{intrinsic}; this condition is trivially met when the evidence is defined in terms of nodes and edges visited \citep{collazo2018chain}. 


Unlike most other graphical models, the observation of a node or edge in a CEG \textit{reduces} the complexity of the process under consideration by rendering some other nodes and edges unvisited with probability one. Conditional on the evidence, such nodes and edges can be removed without loss of information. If we denote a CEG by $\calC$, and evidence by $\scrE$, then the $\scrE$-reduced graph of $\calC$ does not contain the nodes and edges rendered impossible by $\scrE$. The propagation algorithm for CEGs is a backward-forward message-passing algorithm involving closed-form calculations to revise the probabilities associated with the $\scrE$-reduced graph of $\calC$. The technical details of the propagation algorithm are given in Appendix B in the supplementary material.

\section{The \pkg{cegpy} Package} \label{sec:cegpy_package}

\pkg{cegpy} is the first \proglang{Python} implementation of CEGs, the first across all languages to support structurally asymmetric processes and the first to support probability propagation. It is developed using a Bayesian framework which provides a structured way to incorporate prior knowledge, to update the posterior as more data is received, and to perform modelling with a combination of expert knowledge and data when working on problems with limited data. 

\subsection{Python Implementation} \label{subsec:python_implementation}

\begin{figure}[hb]
    \centering
    \includegraphics[width=\textwidth]{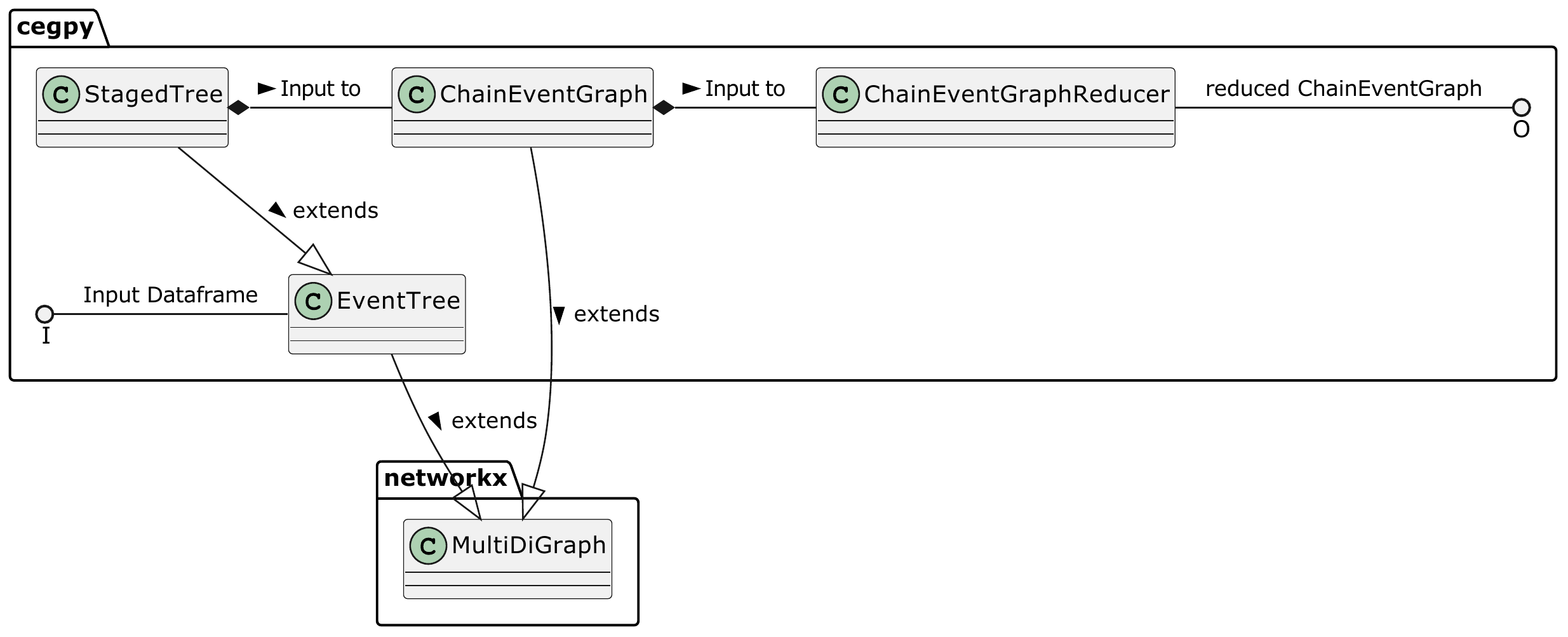}
    \caption{Class diagram of \pkg{cegpy}.}
    \label{fig:class_diagram}
\end{figure}

\pkg{cegpy} is built in \proglang{Python}, and makes use of the open source packages \pkg{pandas}, \pkg{NetworkX}, and \pkg{GraphViz}. It is designed to harness the object-oriented functionalities of \proglang{Python}. The class diagram in Figure \ref{fig:class_diagram} shows the object-oriented inheritance structure of the various classes in the package. The EventTree, StagedTree, ChainEventGraph and ChainEventGraphReducer (corresponding to an $\scrE$-reduced CEG, see Section \ref{subsubsec:probability_propagation}) are all object classes. From the inheritance structure, we can see that, for instance, the StagedTree class inherits the features of the EventTree class and extends it. The EventTree class and ChainEventGraph class inherit from the MultiDiGraph (directed multi-graph) class of \pkg{NetworkX} which is a well-developed and thoroughly tested package for studying graphs and networks in \proglang{Python}. 

The EventTree class is the entry-point for data into the package. It converts the input data into an event tree so that a model selection algorithm can be run on it to then create a StagedTree. To construct the event tree, the EventTree object scans each row of a column-based data set, and counts all the unique sequences of events, i.e. paths, contained in it. This data is stored in a \proglang{Python} dictionary, where each key in the dictionary represents an edge (expressed as a path from the root up to that edge) in the tree, and maps to the number of times that path has been observed in the dataset which corresponds to the number of transitions along that edge. 

\begin{table}[h!]
\footnotesize
\begin{tabular}{l|l|l|l}
\textbf{Housing Assessment}     & \textbf{Risk}      & \textbf{Treatment}                    & \textbf{Fall}       \\
\hline
Community Not Assessed & Low Risk  & Not Referred and Not Treated & Fall       \\
Community Not Assessed & High Risk & Not Referred and Not Treated     & Fall       \\
Community Assessed     & Low Risk  &                      --        & Don't Fall \\
Community Assessed     & High Risk & Referred and Treated         & Fall      
\end{tabular}
\caption{An example dataset illustrating how the falls intervention data is stored.}
\label{tab:example_data}
\end{table}

For example, consider the example dataset for the falls intervention in Table \ref{tab:example_data}. This is transformed into a dictionary like so: 

\begin{minted}[breaklines, fontsize=\footnotesize]{python}
edges = {
    ("Community Assessed", ): 2,
    ("Community Not Assessed", ): 2,
    ("Community Assessed", "High Risk"): 1,
    ("Community Assessed", "Low Risk"): 1,
    ("Community Not Assessed", "High Risk"): 1,
    ("Community Not Assessed", "Low Risk"): 1,
    ("Community Assessed", "High Risk", "Referred and Treated"): 1,
    ("Community Assessed", "Low Risk", "Don't Fall"): 1,
    ("Community Not Assessed", "High Risk", "Not Referred and Not Treated"): 1,
    ("Community Not Assessed", "Low Risk", "Not Referred and Not Treated"): 1,
    ("Community Assessed", "High Risk", "Referred and Treated", "Fall"): 1,
    ("Community Not Assessed", "High Risk", "Not Referred and Not Treated", "Fall"): 1,
    ("Community Not Assessed", "Low Risk", "Not Referred and Not Treated", "Fall"): 1,
}
\end{minted}

By design, the paths are ordered alphabetically, which enables consistent node-naming, even when the rows in the dataset are reordered. Once the paths are determined, the edges dictionary is used to create a \pkg{NetworkX} MultiDiGraph object, which is then used as the data structure representation of the event tree.

As the StagedTree class inherits from the EventTree class, the former is simply a special case of the latter. A StagedTree object determines which nodes of the EventTree are in the same stage and applies a colour scheme to them to show the distinct stages. The \pkg{cegpy} package contains an implementation of the AHC algorithm described in Section \ref{subsubsec:model_selection} for identifying the stages from the EventTree. Under a Bayesian framework, the algorithm is initiated with a prior specification for each situation (i.e. each initial singleton stage) in the EventTree. With the default settings in \pkg{cegpy}, this is done using the approach of \citet{collazo2018chain} whereby an imaginary sample size for the root is set and this is then propagated uniformly through the EventTree. The default imaginary sample size is given by the maximum number of edges emanating out of any situation in the EventTree. Further, \pkg{cegpy} also supports specification of priors for the colouring of the StagedTree (and thus, the structure of the ChainEventGraph) by indicating which situations are allowed to be merged together. This is done by specifying a \textit{hyperstage} \citep{collazo2017thesis} which is a collection of sets such that two situations cannot be in the same stage unless they belong to the same set in the hyperstage. Under a default setting in \pkg{cegpy}, all situations which have the same number of outgoing edges and equivalent set of edge labels are in the same set within the hyperstage. Note that users can specify their own prior, imaginary sample size or hyperstage.

The AHC implementation is run within the StagedTree object and it is parallelised for efficiency. Once the stages have been identified, a ChainEventGraph object can be created by passing the StagedTree as input. The ChainEventGraph object merges the nodes that are in the same position. To do this, it uses the transformation algorithm described in \citet{shenvi2020constructing}. \pkg{cegpy} also includes an implementation of the probability propagation algorithm described in Section \ref{subsubsec:probability_propagation}. In order to propagate probabilities through a CEG, the ChainEventGraphReducer object is first instantiated from the ChainEventGraph object. Evidence $\scrE$ can be specified to this object in terms of the nodes and edges as certain or uncertain evidence. Once all the evidence has been provided, the $\scrE$-reduced CEG with the updated probabilities can then be created as a new ChainEventGraph object.

\subsection{Related Work} \label{subsec:related_work}

There are two previous packages that can learn and visualise CEGs from data. In 2017, the \proglang{R} package \pkg{ceg} \citep{ceg} was the first package able to learn CEGs from data. This package implements the AHC for Bayesian model selection in CEGs. In 2021, the \proglang{R} package \pkg{stagedtrees} \citep{carli2022r} was released, which included several score-based and clustering-based algorithms for non-Bayesian model selection in CEGs such as hill-climbing, backward hill-climbing and k-means. Whilst both these packages are able to represent context-specific conditional independencies, neither are able to represent asymmetric structures, i.e. non-stratified staged trees and CEGs. Therefore, neither of these packages can be used to model processes with asymmetrical unfolding of events such as those described in Section \ref{subsec:asymmetries}; see examples in Appendix C in the supplementary material. \pkg{cegpy} fills this gap. 

\begin{table}[]
\small
\begin{tabular}{c|cccc}
Package & \multicolumn{1}{l}{Language} & \multicolumn{1}{l}{CEG class} & \multicolumn{1}{l}{Model selection} & Propagation \\ \hline
\pkg{ceg}         & \proglang{R} & Stratified only & Bayesian & \xmark \\
\pkg{stagedtrees} & \proglang{R} & Stratified only & Frequentist & \xmark \\
\pkg{cegpy}       & \proglang{Python} & \shortstack{Stratified and \\ non-stratified}  & Bayesian & \cmark
\end{tabular}
\caption{A comparison of the three packages available for modelling with CEGs.}
\label{tab:comparison}
\end{table}

The path-based approach of \pkg{cegpy} is in contrast to the column-based approach of other CEG packages such as \pkg{ceg} and \pkg{stagedtrees}. In the former approach, all the data is associated with edges of the event tree and it corresponds to using events as the building blocks of the model. It can routinely handle non-stratified CEGs. On the other hand, the column-based approach associates the data to the variables of the model and to their corresponding state spaces. This approach makes it extremely difficult to model non-stratified CEGs. For instance, the structural missing values associated with the \textit{Treatment} variable for individuals who are not assessed cannot be recorded easily within the column-based approach. An alternative is to create a new category for the \textit{Treatment} variable called ``Not referred \& not treated'' which then renders the counts on the other categories of this variable into structural zeros leading to redundancies in the parameters and loss of representation. 

Finally, \proglang{Python}'s object-oriented architecture lends itself well to extensibility. The functionality and CEG classes supported by \pkg{cegpy} can be easily built upon using inheritance. For example, to create a new class of CEGs in \pkg{cegpy} with arbitrary holding time distributions (such as in \citet{barclay2015dynamic}), a new TemporalEventTree class can be created which inherits from the EventTree class and extends it to handle the holding times in the input dataset. Similarly, a TemporalStagedTree class can be created such that it inherits the initialisation and functions from the TemporalEventTree class as well as just the functions from the StagedTree class.

\section{An Illustrative Example} \label{sec:worked_example}

In this section, we illustrate the key functionalities of the \pkg{cegpy} package through the analysis of a structurally asymmetric process. We revisit the public health intervention to reduce falls-related injuries and fatalities among the elderly as described in Example \ref{ex:falls_description}. We use the synthetic dataset simulated by \citet{shenvi2018modelling} and provided with the supplementary materials. Note that illustrations and guidance for the full range of functionalities support by \pkg{cegpy} can be found at {\footnotesize \url{https://github.com/peterrhysstrong/cegpy-binder}.}

\subsection{Creating the Event Tree}
The Falls dataset provides information concerning adults over the age of 65, and includes four categorical variables as given below with their state spaces:
\begin{itemize}[itemsep=0pt, topsep=0.5em]
    \item Living situation and whether they have been assessed, state space: \{Communal Assessed, Communal Not Assessed, Community Assessed, Community Not Assessed\}; 
    \item Risk of a future fall, state space: \{High Risk, Low Risk\};
    \item Referral and treatment status, state space: \{Not Referred \& Not Treated, Not Referred \& Treated, Referred \& Treated\};
    \item Outcome, state space: \{Fall, Don't Fall\}.
\end{itemize}
\noindent Recall from the description in Example \ref{ex:falls_description} that this process has structural asymmetries. None of the individuals assessed to be low risk are referred to the falls clinic and thus, for this group, the count associated with the `Referred \& Treated' category is a structural zero. Moreover, for individuals who are not assessed, their responses are structurally missing for the referral and treatment variable. 

Observe that since \pkg{cegpy} constructs the event tree by creating a dictionary of the paths in the input dataset, there is no need to specify structural zeros as they do not occur in the dataset. On the other hand, we encode structural missing values in the dataset as NaNs. For example, a NaN value in the column relating to the referral and treatment variable is interpreted by \pkg{cegpy} as a structural missing value. 

\begin{minted}[breaklines, fontsize=\footnotesize]{python}
from cegpy import EventTree
import pandas as pd

df = pd.read_excel('data/Falls_Data.xlsx')
print(df.head(5))
\end{minted}
\footnotesize
\begin{verbatim}
output:
        HousingAssessment       Risk Treatment        Fall
0  Community Not Assessed   Low Risk       NaN        Fall
1  Community Not Assessed  High Risk       NaN        Fall
2  Community Not Assessed   Low Risk       NaN  Don't Fall
3  Community Not Assessed   Low Risk       NaN  Don't Fall
4  Community Not Assessed   Low Risk       NaN        Fall
\end{verbatim}
\normalsize

\begin{figure}[!h]
    \centering
        \includegraphics[width =0.65\textwidth] {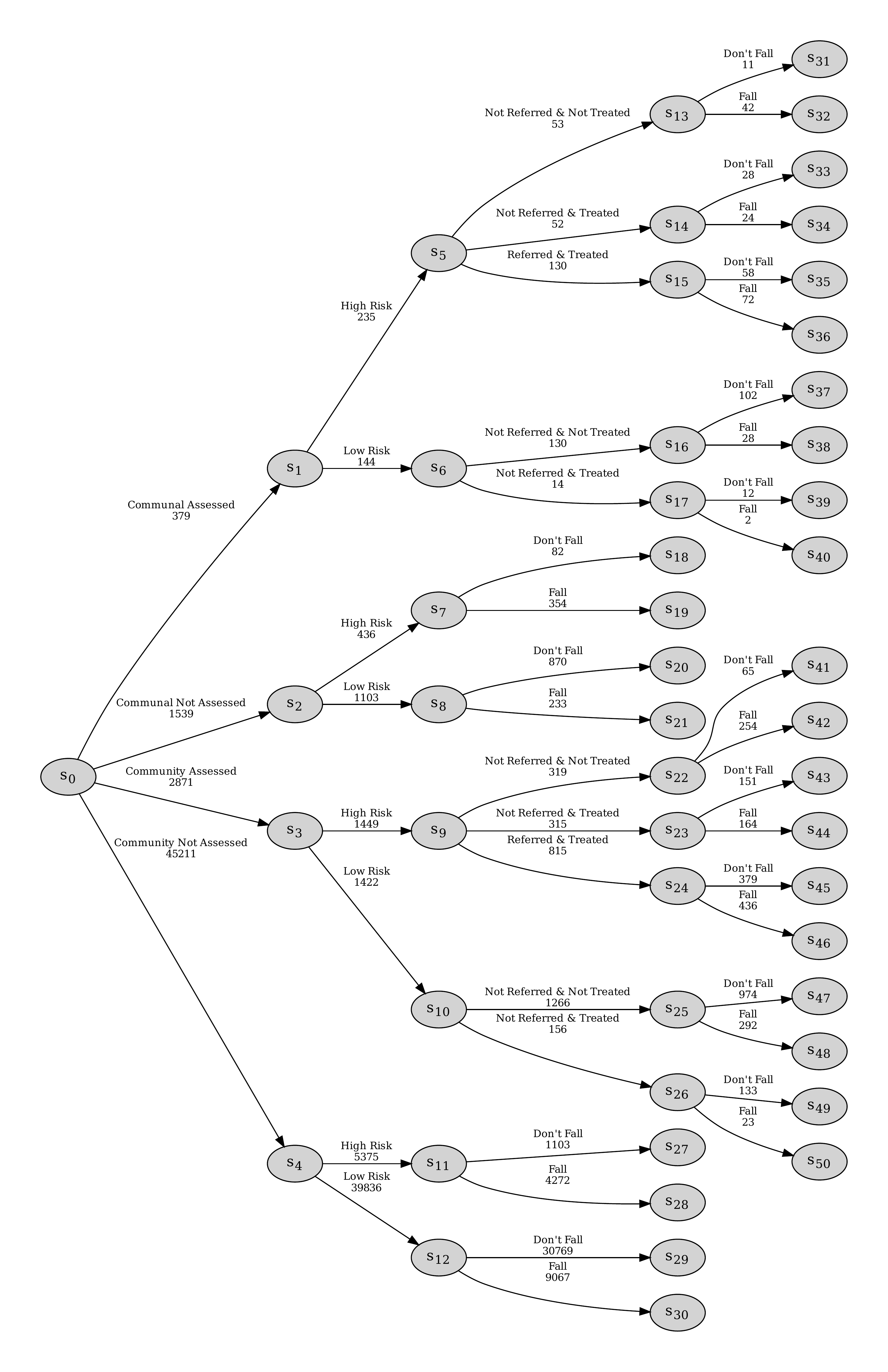}
        \caption{Event tree output for the falls dataset.}
    \label{fig:example_falls_event_tree}
\end{figure}

The event tree can be constructed from the falls dataset by initialising an EventTree object using the code given below and as shown in Figure \ref{fig:example_falls_event_tree}.

\begin{minted}[breaklines, fontsize=\footnotesize]{python}
et = EventTree(df)
et.create_figure()
\end{minted}

Note here that any paths that should logically be in the event tree description of the process but are absent from the dataset due to sampling limitations would need to be manually added by the user using the \mbox{\textit{sampling\_zero\_paths}} argument when initialising the EventTree object. Further, not all missing values in the dataset will be structurally missing. To demarcate the difference, a user can give different labels to the structural and sampling missing values in the dataset and provide these labels to the \textit{struct\_missing\_label} and \textit{missing\_label} arguments respectively when initialising the EventTree object.

\subsection{Creating the Staged Tree}

We now look at creating a staged tree from our dataset. To do this, we initialise a StagedTree object with our dataset as the input. Note that it is not necessary to first initialise an EventTree object. To create a staged tree, we must first identify the stages in the event tree. We do this by running the AHC algorithm within the StagedTree object. The code and output below show the default settings of the hyperstage, alpha (imaginary sample size at the root, see Section \ref{subsec:python_implementation}) and prior for the falls dataset. The priors and posteriors are saved as fractions to maintain accuracy through the iterative calculations.


\begin{minted}[breaklines, fontsize=\footnotesize]{python}
from cegpy import StagedTree
st = StagedTree(df)
print('default hyperstage:',st._create_default_hyperstage())
print('default alpha:',st._calculate_default_alpha())
print('default prior:',
        st._create_default_prior(st._calculate_default_alpha()))

Output:
default hyperstage: [['s0'], ['s1', 's2', 's3', 's4'], ['s5', 's9'],
['s6', 's10'],['s7', 's8', 's11', 's12', 's13', 's14', 's15', 's16', 
's17', 's22', 's23', 's24', 's25', 's26']]
default alpha: 4
default prior: [[Fraction(1, 1), Fraction(1, 1), Fraction(1, 1),
Fraction(1, 1)],[Fraction(1, 2), Fraction(1, 2)], [Fraction(1, 2),
Fraction(1, 2)],[Fraction(1, 2), Fraction(1, 2)], [Fraction(1, 2),
Fraction(1, 2)],[Fraction(1, 6), Fraction(1, 6), Fraction(1, 6)],
[Fraction(1, 4),Fraction(1, 4)], [Fraction(1, 4), Fraction(1, 4)],
[Fraction(1, 4),Fraction(1, 4)], [Fraction(1, 6), Fraction(1, 6),
Fraction(1, 6)],[Fraction(1, 4), Fraction(1, 4)], [Fraction(1, 4),
Fraction(1, 4)],[Fraction(1, 4), Fraction(1, 4)], [Fraction(1, 12),
Fraction(1, 12)],[Fraction(1, 12), Fraction(1, 12)], [Fraction(1, 12),
Fraction(1, 12)],[Fraction(1, 8), Fraction(1, 8)], [Fraction(1, 8),
Fraction(1, 8)],[Fraction(1, 12), Fraction(1, 12)], [Fraction(1, 12),
Fraction(1, 12)],[Fraction(1, 12), Fraction(1, 12)], [Fraction(1, 8),
Fraction(1, 8)],[Fraction(1, 8), Fraction(1, 8)]]
\end{minted}

\begin{figure}[!h]
    \centering
        \includegraphics[width =0.65\textwidth]{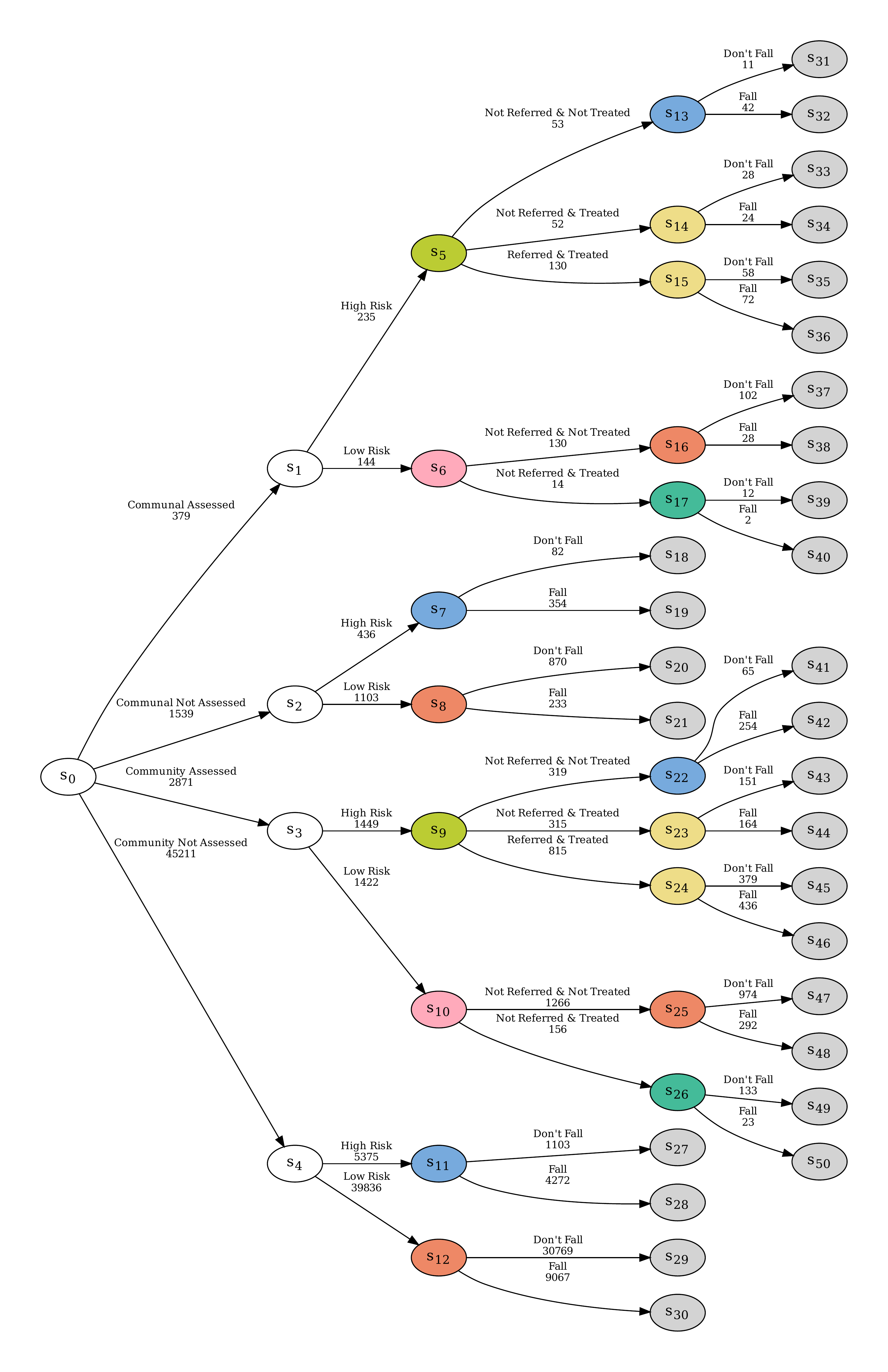}
        \caption{Staged tree output for the falls dataset with default priors.}
    \label{fig:example_falls_staged_tree}
\end{figure}

We now run the AHC algorithm with the above default settings using the code below and generate the associated staged tree figure as shown in Figure \ref{fig:example_falls_staged_tree}. Additionally, a user can specify a list of colours or palette to be used in the staged tree and its corresponding CEG. In this example, we have used a colourblind-friendly palette as shown by the \textit{colours} list below.
\begin{minted}[breaklines, fontsize=\footnotesize]{python}
colours = ['#BBCC33','#77AADD','#EE8866','#EEDD88','#FFAABB','#44BB99']
st.calculate_AHC_transitions(colour_list=colours)
st.create_figure()
\end{minted}

\subsection{Creating the Chain Event Graph}

\begin{figure}[!h]
    \centering
        \includegraphics[width =0.9\textwidth]{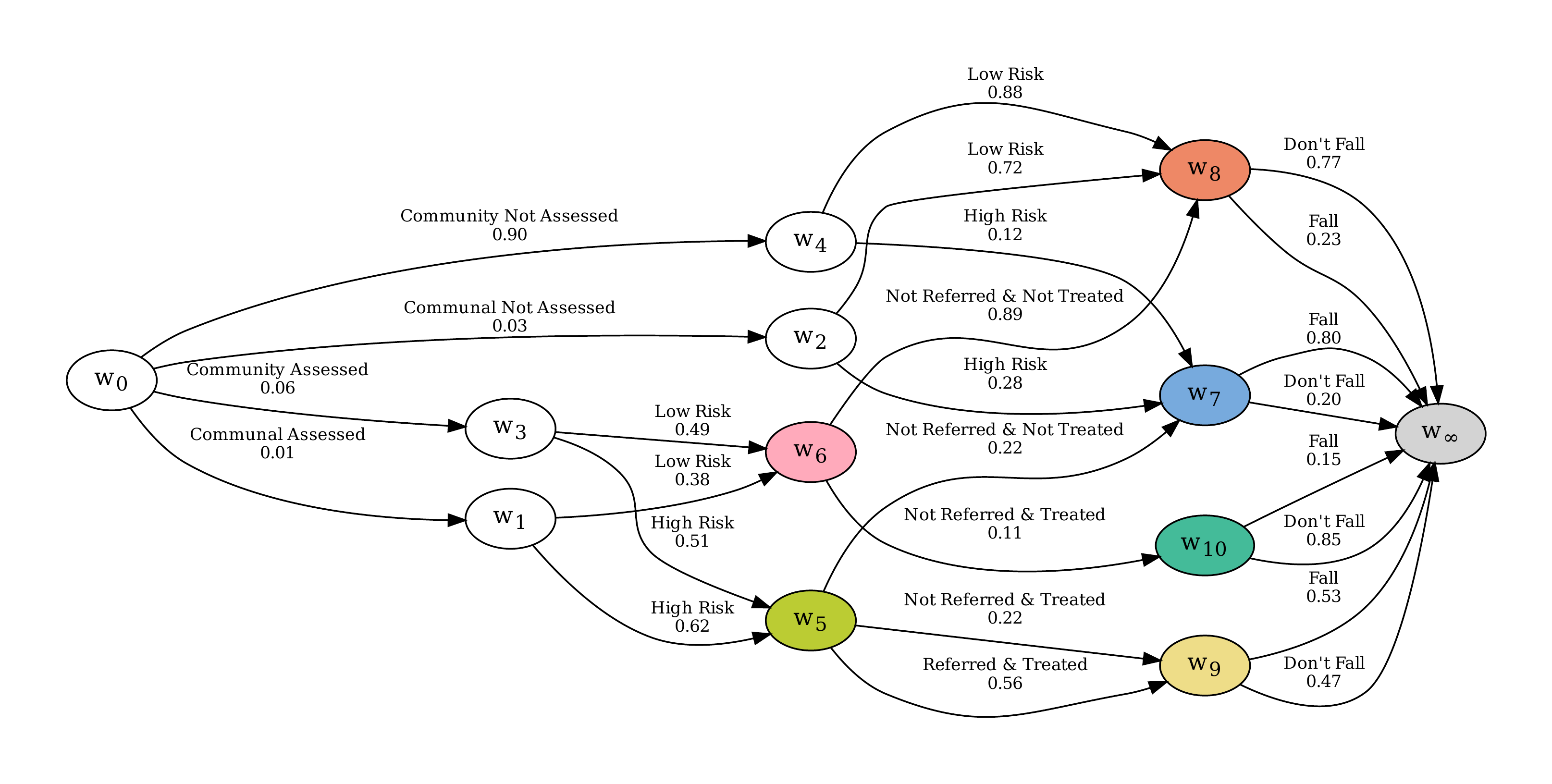}
        \caption{CEG output for the falls dataset.}
    \label{fig:example_falls_ceg}
\end{figure}

Once the stages have been identified by running the AHC algorithm on the StagedTree object, we can initialise a ChainEventGraph object that takes the StagedTree object as an input. Using this StagedTree object, the ChainEventGraph object can generate the CEG figure using the code below and as shown in Figure \ref{fig:example_falls_ceg}.

\begin{minted}[breaklines, fontsize=\footnotesize]{python}
from cegpy import ChainEventGraph
ceg = ChainEventGraph(st)
ceg.create_figure()
\end{minted}

\subsection{Probability Propagation on the Chain Event Graph}

\begin{figure}[!h]
    \centering
        \includegraphics[width =0.9\textwidth]{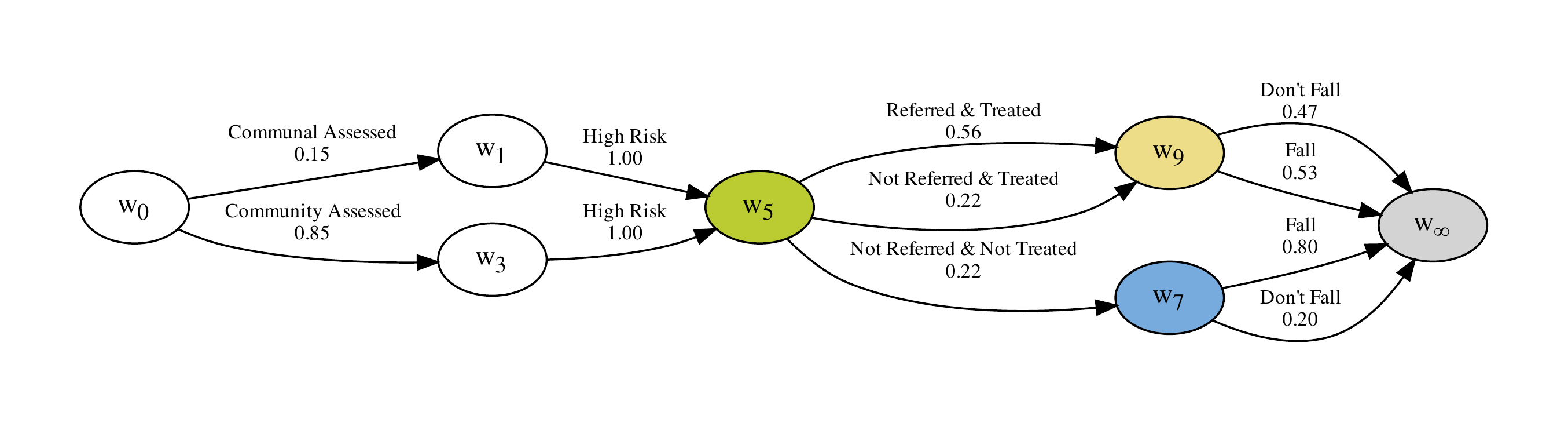}
        \caption{CEG output for the falls dataset after propagating the observation of node $w_5$, i.e. individuals who are assessed and are high risk.}
    \label{fig:example_falls_tceg}
\end{figure}

Finally, we demonstrate how \pkg{cegpy} can be used for probability propagation on a given CEG after observing some evidence associated with it. For instance, suppose that we have observed an assessed, high risk individual. This is equivalent to observing the node $w_5$ in the CEG in Figure \ref{fig:example_falls_ceg} with certainty. The probabilities associated with the CEG can be updated in light of this certain observation by first initialising a ChainEventGraphReducer object with the CEG as the input and then adding the certain evidence as shown in the code below. The graph and conditional probabilities of the updated CEG are given in Figure \ref{fig:example_falls_tceg}. We can see that based on this observation, the probability that the observed individual is from a communal establishment is updated from $0.04$ (sum of communal assessed and communal not assessed) to $0.15$.

\begin{minted}[breaklines, fontsize=\footnotesize]{python}
from cegpy import ChainEventGraphReducer
rceg = ChainEventGraphReducer(ceg)
rceg.add_certain_node('w5')
rceg.graph.create_figure()
\end{minted}

\begin{figure}[!h]
    \centering
        \includegraphics[width =0.9\textwidth]{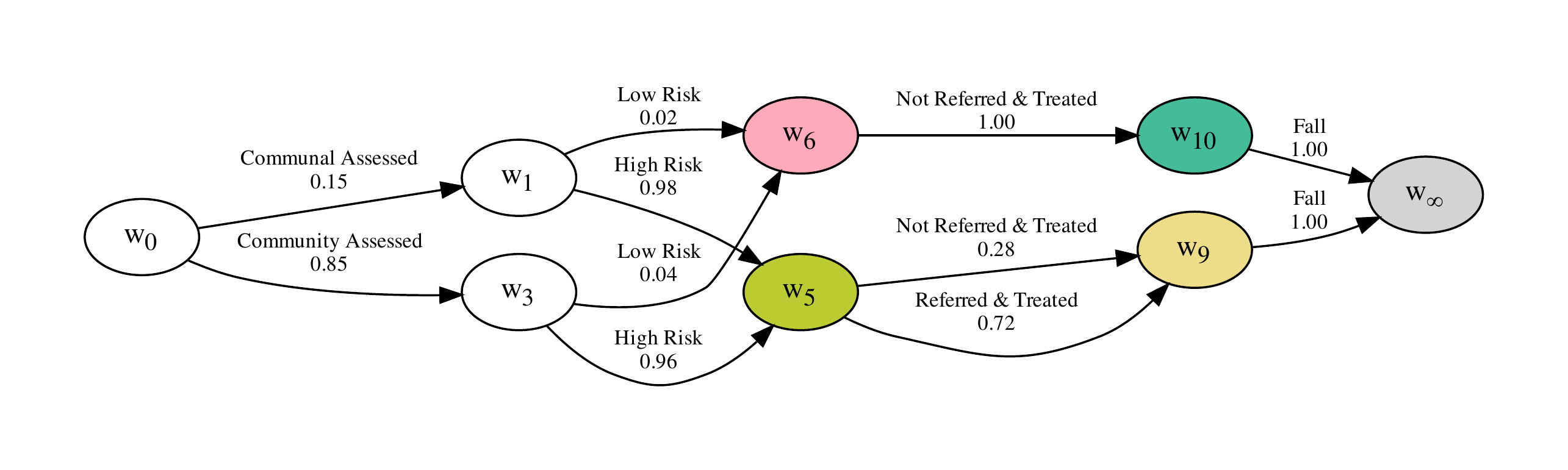}
        \caption{CEG output for the falls dataset after propagating the uncertain evidence over the nodes: $w_9$, $w_{10}$ and over the edges: ($w_9$, $w_\infty$, Fall), ($w_{10}$, $w_\infty$, Fall).}
    \label{fig:example_tceg_uncertain}
\end{figure}

We can also use \pkg{cegpy} to propagate uncertain evidence. For example, consider the evidence that the individuals under consideration had been treated but they still suffered a fall. Based on this information, we know that the individuals must have passed through one of the following sequences of edges in the CEG in Figure \ref{fig:example_falls_ceg}: (i) $(w_5, w_{9}, \text{`Not Referred \& Treated'})$ and $(w_9, w_\infty, \text{`Fall'})$; (ii) $(w_5, w_9, \text{`Referred \& Treated'})$ and  $(w_9, w_\infty, \text{`Fall'})$; or (iii) $(w_6, w_{10}, \text{`Not Referred \& Treated'})$ and $(w_{10}, w_\infty, \text{`Fall'})$. To simplify, this is equivalent to having uncertain evidence about nodes $w_9$ and $w_{10}$, and about the edges $(w_9, w_\infty, \text{`Fall'})$ and $(w_{10}, w_\infty, \text{`Fall'})$. As earlier, to update the graph and the conditional probabilities in the CEG, we can initialise a ChainEventGraphReducer object with the CEG as the input, add both sets of uncertain evidence and obtain the updated figure. The code for this is given below and the graph of the updated CEG is in Figure \ref{fig:example_tceg_uncertain}.

\begin{minted}[breaklines, fontsize=\footnotesize]{python}
from cegpy import ChainEventGraphReducer
rceg = ChainEventGraphReducer(ceg)
rceg.add_uncertain_node_set({"w9", "w10"})
rceg.add_uncertain_edge_set_list([{('w9',ceg.sink, 'Fall'),
                                ('w10', ceg.sink, 'Fall')}])
rceg.graph.create_figure()
\end{minted}

\section{Discussion} \label{sec:discussion}


\pkg{cegpy} is an open-source \proglang{Python} package that facilitates modelling with staged trees and CEGs, providing functionality for Bayesian model selection and probability propagation. This package is the first implementation of staged trees and CEGs in \proglang{Python}, and, unlike previous implementations in \proglang{R} which focus only on the stratified class, \pkg{cegpy}’s functionality extends to all staged trees and CEGs, stratified and non-stratified alike. Further, it is the first package that provides support for probability propagation. Therefore, \pkg{cegpy} can support users with categorical data to create models of processes with structural asymmetries, which can be analysed to understand complex dependence structures. We discuss below a few avenues for greatly enhancing the current functionality provided by \pkg{cegpy}\footnote{Contributions on \url{https://github.com/g-walley/cegpy} are always welcome and will be much appreciated.}.
 
In the current version of \pkg{cegpy}, we have focused on Bayesian methods. However, it is straightforward to implement other classical methods and we plan to do this in a future version. Moreover, currently \pkg{cegpy} only provides support for the AHC algorithm. Other existing Bayesian model selection algorithms have considerable drawbacks: dynamic programming \cite{silander2013dynamic} is computationally infeasible for all but the smallest of data sets. Further research is needed to explore computationally efficient Bayesian model selection techniques for CEGs. \citet{strong2022bayesian}'s work on Bayesian model averaging using a modification of the AHC algorithm has been implemented as an extension to \pkg{cegpy} (see {\footnotesize \url{https://github.com/peterrhysstrong/cegpy_BMA}}) and we plan to make this available in a future version of the package. Further extensions could implement other model selection techniques, such as those in \pkg{stagedtree}. 

As described in Section \ref{subsec:related_work}, \pkg{cegpy} uses a path-based approach to constructing the event tree. Thereby, sampling zeros paths are not automatically filled in for unobserved combinations of variables. Currently, users must add these paths manually. Filling in of sampling zero paths can be automated by assuming that the tree is stratified. This can be added as an argument for the EventTree class to create these paths at the point of initialisation of an EventTree object.

Finally, for the purposes of expert elicitation, it would be extremely useful to enable a user to directly specify an event tree, staged tree or CEG structure -- with colouring and possibly, with parameters -- in the \pkg{cegpy} package. Of course, model selection algorithms cannot be used due to the absence of data but it would be beneficial for visualisation and evidence propagation. We are currently looking into adding this functionality by directly importing graphs specified using the DOT language used by \pkg{GraphViz}.


\bibliographystyle{elsarticle-harv} 
\bibliography{biblio}

\end{document}


\begin{frontmatter}



\title{Supplementary Material for:\\
cegpy: Modelling with Chain Event Graphs in Python}


\author[inst1]{Gareth Walley\fnref{label1}}

\author[inst2]{Aditi Shenvi\corref{cor1}\fnref{label1}}
\ead{aditi.shenvi@gmail.com}

\author[inst3,inst4]{Peter Strong}

\author[inst2]{Katarzyna Kobalczyk}

\cortext[cor1]{Corresponding author.}

\fntext[label1]{Both authors contributed equally to this research.}

\affiliation[inst1]{
            city={Kenilworth},
            postcode={CV8 1JY}, 
            country={UK}}
\affiliation[inst2]{organization={Statistics Department, University of Warwick},
            city={Coventry},
            postcode={CV4 7AL}, 
            country={UK}}
\affiliation[inst3]{organization={Centre for Complexity Science, University of Warwick},
            city={Coventry},
            postcode={CV4 7AL}, 
            country={UK}}
            
\affiliation[inst4]{organization={Alan Turing Institute},
            city={London},
            postcode={NW1 2DB}, 
            country={UK}}
            
\end{frontmatter}

\renewcommand{\thesection}{\Alph{section}}

\section{AHC Algorithm} \label{app:ahc}

Consider a staged tree where $\mathbb{U} =$ $\{u_1, u_2, \ldots, u_k\}$. Let stage $u_i$ for $i = 1, 2, \ldots, k$ have $k_i$ emanating edges and let its conditional transition parameter vector be given by $\pmb{\theta}_i = \{\theta_{i1}, \theta_{i2}, \ldots, \theta_{ik_i}\}$ where $\theta_{ij}$ is the probability that an individual in some situation $s \in u_i$ traverses along its $j$th outgoing edge, for $j \in \{1,2, \ldots, k_i\}$. For each $\pmb{\theta}_i$, we set a Dirichlet prior distribution with parameter vector $\pmb{\alpha}_i = (\alpha_{i1}, \alpha_{i2}, \ldots, \alpha_{ik_i})$ where $\alpha_{ij} > 0, j \in \{1, 2, \ldots, k_i\}$. Suppose that we have a complete ancestral sample given by $\textbf{y} = \{y_1, y_2, \ldots, y_{k}\}$ such that each $y_{i} = (y_{i1}, y_{i2}, \ldots, y_{ik_i})$ is a vector summarising the number of individuals $y_{ij}$, $j \in \{1,2, \ldots, k_i\}$ that start in some situation $u_{i}$ and traverse along its $j$th edge.

Under the standard assumptions of complete ancestral sampling, local and global parameter independence, the posterior distribution of the $\pmb{\theta}_i$ also follows a Dirichlet distribution with parameter vector ${\pmb{\alpha}^*_i = (\alpha_{i1}^*, \alpha_{i2}^*, \ldots, \alpha_{ik_i}^*)}$ where $\alpha_{ij}^* = \alpha_{ij} + y_{ij}, j \in \{1, 2, \ldots, k_i\}$, $i \in \{1,2, \ldots, k\}$. If each staged tree model is assumed to be equally likely \textit{a priori}, then its log marginal likelihood score $Q(\calS)$ can be obtained as follows
\begin{equation}
Q(\calS) = \sum_{i = 1}^{k} \bigg \{ g(\bar{\pmb{\alpha}}_i) - g(\bar{\pmb{\alpha}}_i^*) + \sum_{j=1}^{k_i} \{ g(\alpha_{ij}^*) - g(\alpha_{ij}) \} \bigg \}
\label{eq:score}
\end{equation}
\noindent where $g(\cdot) = \log \Gamma(\cdot)$, $\bar{\textbf{v}} = \sum_{i=1}^{n} v_{i}$ for any vector $\textbf{v} = (v_1, v_2, \ldots, v_n)$, and $\Gamma(z) = \int_{0}^{\infty} x^{z-1}\exp(-x) \, dx$ is the Gamma function.

The AHC algorithm is a local greedy search algorithm which uses a bottom-up hierarchical clustering methodology beginning with the coarsest clustering treating each situation as a singleton cluster and successively merging pairs of stages until the log marginal likelihood score cannot be improved further. The comparison of log marginal likelihood scores of two staged trees $\calS$ and $\calS'$ is given by the log Bayes factor \citep{kass1995bayes} as follows
\begin{align}
\log BF (\calS', \calS) &= Q(\calS') - Q(\calS).  
\label{eq:Bayes_factor}
\end{align} 

Since the AHC does pairwise comparisons, it essentially compares one-nested models. In this case, the $\log BF$ calculation can be simplified as follows. Two staged trees $\calS$ and $\calS'$ are said to be \textit{one-nested} when two stages $u_i, u_j \in \mathbb{U}_\calS$ in $\calS$ are represented by a single stage $u_{i \oplus j} \in \mathbb{U}_{\calS'}$ in $\calS'$. Thus, the situations in $u_i, u_j$ and $u_{i \oplus j}$ have the same number of outgoing edges denoted here by $k$. The $\log BF$ is then a linear combination of the terms involving the hyperparameters associated with stages $u_i, u_j$ and $u_{i \oplus j}$ only as given by
\begin{align}
\log BF (\calS', \calS) &= g(\bar{\pmb{\alpha}}_{i \oplus j}) - g(\bar{\pmb{\alpha}}_i) - g(\bar{\pmb{\alpha}}_j) - g(\bar{\pmb{\alpha}}_{i \oplus j} ^*) + g(\bar{\pmb{\alpha}}_i^*) + g(\bar{\pmb{\alpha}}_j^*)  \nonumber \\
& \,\, + \sum_{l = 1}^k \{g(\alpha_{i \oplus j,l}^*) - g(\alpha_{il}^*)  - g(\alpha_{jl}^*) -  g(\alpha_{i \oplus j, l})  + g(\alpha_{il}) + g(\alpha_{jl})  \}.
\label{eq:one_nested_Bayes_factor}
\end{align} 

The pseudo-code for the AHC algorithm is given in Algorithm \ref{alg:AHC_pseudocode}. In \pkg{cegpy}, the AHC implementation is parallelised by combining steps 7 and 8 such that the $\log BF$ scores of all pairwise merges are calculated in parallel rather than within a \textit{for} loop.

\begin{algorithm}[h!]
    \SetAlgoLined
    \Input{Event tree $\calT$, data $\textbf{y}$, root imaginary sample size $\bar{\pmb{\alpha}_0}$.}
    \Output{Collection of stages $\mathbb{U}$, log marginal likelihood score of the \textit{maximum a posteriori} (MAP) staged tree $\calS$ or equivalently, MAP CEG $\calC$ found by the algorithm.}
    Initialise an array \textit{stages} of each situation $s_i$ in $\calT$.\\
    Initialise an array \textit{data} of $y_i$ for each situation $s_i$ in $\calT$ obtained from $\textbf{y}$.\\
    Initialise an array \textit{priors} of $\pmb{\alpha}_i$ for each situation $s_i$ in $\calT$ obtained by the mass conservation property from $\bar{\pmb{\alpha}}_0$.\\
    Set \textit{score} as the log marginal likelihood score given in Equation \ref{eq:score}.\\
    Set $indicator \leftarrow 1$.\\
    \While{$indicator \neq 0$}{
    \For {every pair of stages in \textit{stages} with same number of outgoing edges and equivalent set of edge labels}{
    Calculate the $\log BF$ as given in Equation \ref{eq:one_nested_Bayes_factor} comparing the structures which merge them together into one stage and keep them apart respectively, all other stages being equal.\\
    \If {no such pair exists}{
    $indicator \leftarrow 0$
    }
    }
    \For {pair $u_i$ and $u_j$ with the largest $\log BF$ score}{
    \If {$\log BF(u_i, u_j )> 0$}{
    $score \leftarrow score +  \log BF(u_i, u_j)$\\
    Update \textit{stages} to add stage $u_{i \oplus j}$ and remove stages $u_i$ and $u_j$.\\
    Update \textit{data} to add $y_{i \oplus j} = y_i + y_j$ and remove $y_i$ and $y_j$.\\
    Update \textit{priors} to add $\pmb{\alpha}_{i \oplus j} = \pmb{\alpha}_i + \pmb{\alpha}_j$ and remove $\pmb{\alpha}_i$ and $\pmb{\alpha}_j$.
    }
    \Else{
    $indicator \leftarrow 0$
    }
    }
    }
    \KwRet{\textit{stages}, \textit{score}} 
\caption{AHC algorithm}
\label{alg:AHC_pseudocode}
\end{algorithm}

\section{Propagation Algorithm} \label{app:prop}

Consider a CEG $\calC$ and intrinsic evidence $\scrE$. Denote the probability of occupying a node $w_i \in V(\calC)$ by $p(w_i)$ and the probability of traversing an edge $e_{ij} = (w_i, w_j, l)$ by $p(e_{ij})$. Let $V^{-1}(s_i)$ denote the nodes whose emanating edges terminate in $s_i$, and $E^{-1}(s_i)$ denote the edges terminating in $s_i$ in the CEG $\calC$. Here $p(.)$ refers to probabilities in $\calC$ and $\hat{p}(.)$ to the updated probabilities in its $\scrE$-reduced CEG. The pseudo-code for the two-pass backward-forward algorithm for propagating $\scrE$ through $\calC$ is given in Algorithm \ref{alg:CEG_propagation}.

\begin{algorithm}[h!]
    \SetAlgoLined
    \Input{Conditional transition probabilities and the $\scrE$-reduced graph for a CEG $\calC$ and intrinsic evidence $\scrE$.}
    \Output{Updated conditional transition probabilities.}
    Set $A \leftarrow \emptyset$, $B \leftarrow \{w_\infty\}$, $\Phi(w_\infty) \leftarrow  1$.\\
    \While{$B \neq \{w_0\}$ (the root node)}{
    \For{$w_j \in B$}{
    \For{$w_i \in V^{-1}(w_j)$}{
    \For {$e_{ij} \in E(w_i) \cap E^{-1}(w_j)$}{
    \If{$e_{ij} \in \Lambda(\scrE)$}{
    $\tau_{e_{ij}} \leftarrow p(e_{ij}) \cdot \Phi(w_j)$}
    \Else{$\tau_{e_{ij}} \leftarrow 0$}
    $A \leftarrow A \cup \{e_{ij}\}$}
    \If{$E(w_i) \subset A$}{
    $\Phi(w_i) = \textstyle \sum_{e_{ij} \in E(w_i)} \tau_{e_{ij}}$ \\
    $B \leftarrow B \cup \{w_i\}$}
    }
    $B \leftarrow B \backslash \{w_j\}$
    }
    }
    \For{$w_i \in V(\calC)$}{
    \For{$e_{i, \cdot} \in E(w_i) \cap \Lambda(\scrE)$}{
    $\hat{p}(e_{i, \cdot}) = \displaystyle\frac{\tau_{e_{i, \cdot}} }{\Phi(w_i)}$
    }
    \For{$e_{i, \cdot} \in E(w_i) \backslash \Lambda(\scrE)$}{
    $\hat{p}(e_{i, \cdot}) = 0$}
    }
    \KwRet{Updated conditional transition probabilities $\hat{p}(.)$}
    \caption{CEG propagation algorithm}
\label{alg:CEG_propagation}
\end{algorithm}

The propagation algorithm uses point estimates of the conditional transition probabilities. In practice, the conditional transition probabilities are unknown. Instead, we use the posterior means of the Dirichlet distributions in the propagation algorithm. \citet{collazo2018chain} showed that the resultant point estimates obtained from the algorithm are equivalent to the means of the updated Dirichlet distributions of the conditional transition parameters.

\section{A Comparison with \pkg{stagedtree} and \pkg{ceg} Packages} \label{app:comparison}
Here we demonstrate how the existing packages would deal with the falls dataset.

\subsection{\pkg{ceg}}
\begin{figure}[h!]
    \centering
        \includegraphics[width =0.7\textwidth]{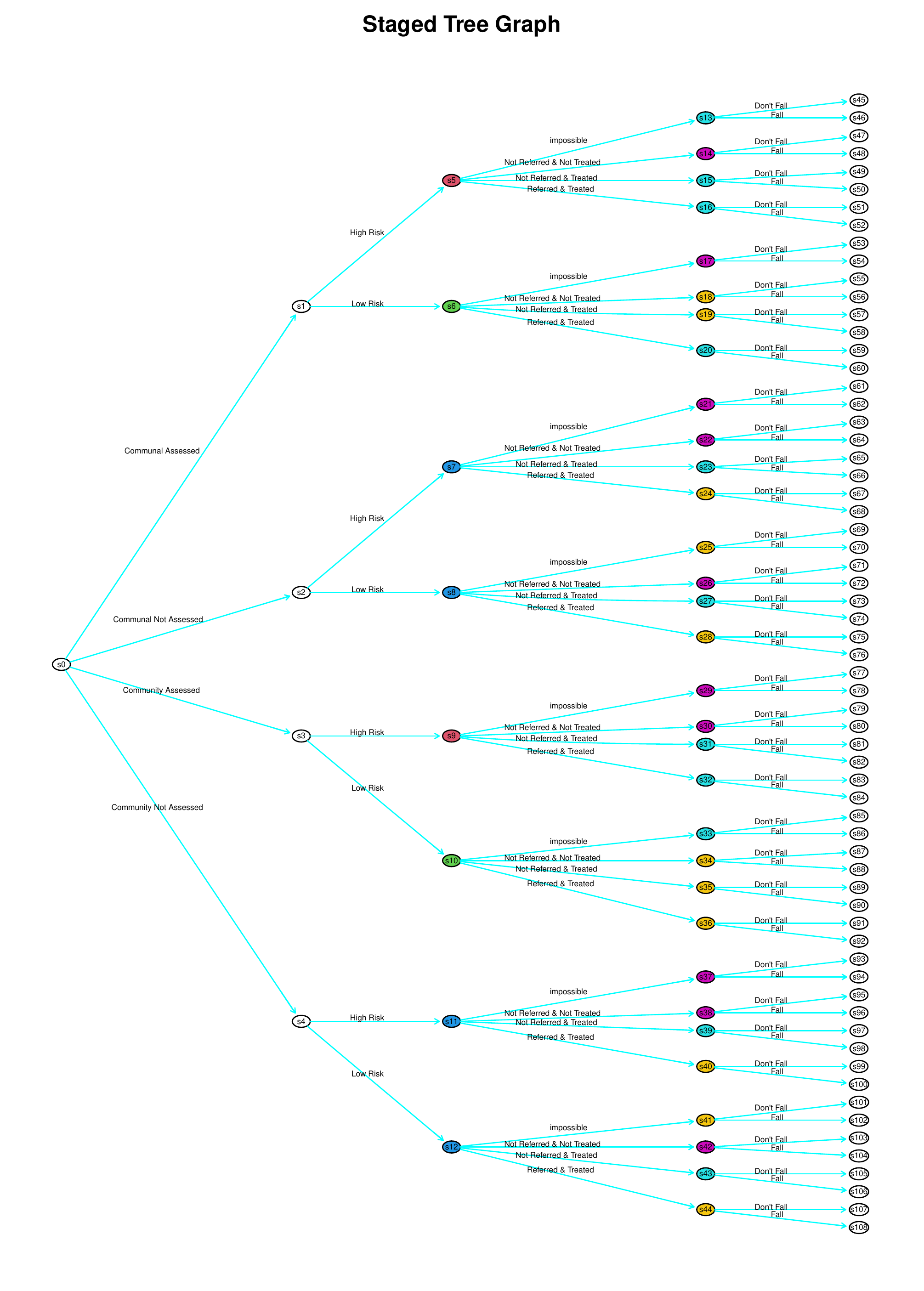}
        \caption{Staged tree for the falls dataset obtained using the \pkg{ceg} \proglang{R} package.}
    \label{fig:ceg_package_falls}
\end{figure}

The \pkg{ceg} package is unable to deal with missing values. Therefore the only way to use this package for the falls dataset is to remove the missing values. To do this, we use the CheckAndCleanData function as described in the documentation \citep{ceg}, and then run the AHC algorithm implemented in \pkg{ceg} to obtain the staged tree in Figure \ref{fig:ceg_package_falls}. This treats the dataset as stratified and is therefore setting a prior over the edges with no observed counts.

\subsection{\pkg{stagedtrees}}

\begin{figure}[h!]
\centering
\includegraphics[width =0.85\textwidth]{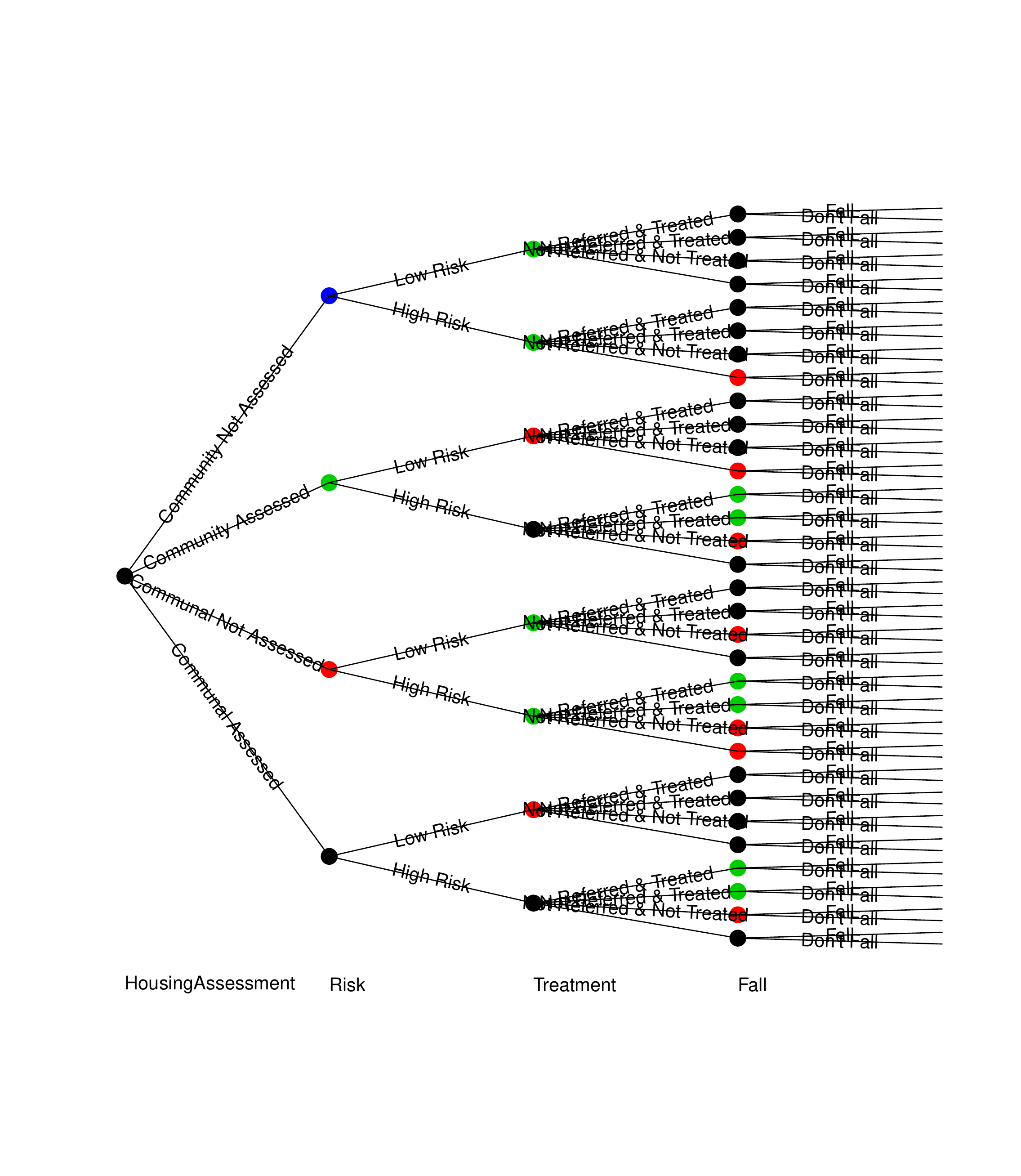} 
\caption{Stratified staged tree for the falls dataset obtained using the \pkg{stagedtrees} \proglang{R} package.}
\label{st_pack}
\end{figure}

The \pkg{stagedtrees} package can deal with missing values. This is done by creating an unlabelled missing outcome on the root to leaf paths. The staged tree obtained by running the backward hill-climbing algorithm with the BIC score as implemented in \pkg{stagedtrees} is given in Figure \ref{st_pack}. This again treats the data as stratified.

\begin{figure}[h]
\centering
    \includegraphics[width =0.85\textwidth]{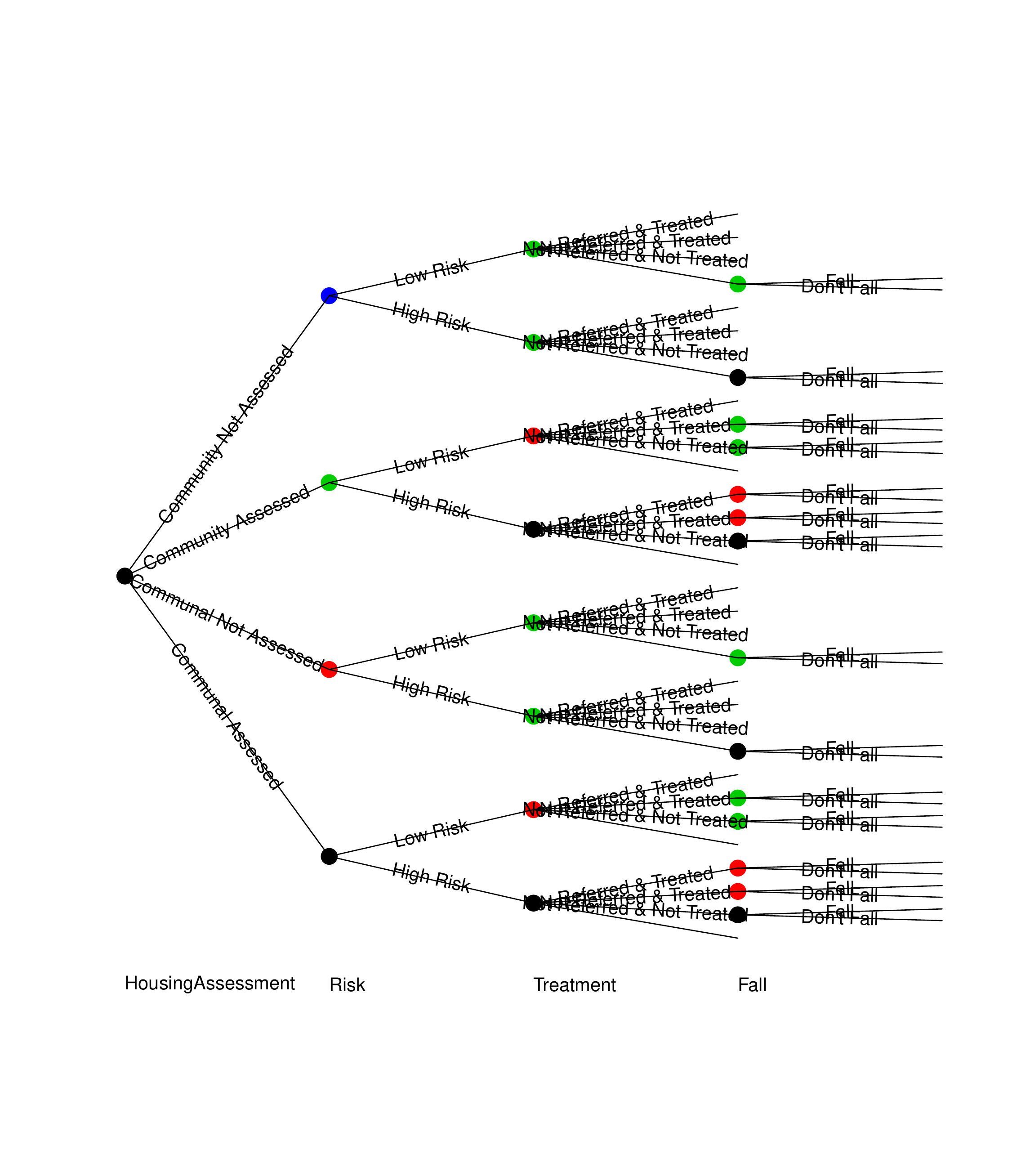}
\caption{Non-stratified staged tree for the falls dataset obtained using the \pkg{stagedtrees} \proglang{R} package. Observe that this does not appropriately represent the falls process.}
\label{st_pack_join}
\end{figure}

An alternative way of treating missing values exists in \pkg{stagedtrees} using the \textit{join\_unobserved=TRUE} argument. By default, this argument is set to FALSE as in Figure \ref{st_pack}. Setting this argument to TRUE merges any situations for which there are no incoming counts. Whilst this results in a non-stratified staged tree, it is not the one that represents the falls process. The resultant staged tree obtained by using the above argument with the backward hill-climbing algorithm and the BIC score is given in Figure \ref{st_pack_join}.

\newpage 
\bibliographystyle{elsarticle-harv} 
\bibliography{biblio}